\documentclass[graybox,vecphys]{svmult}

\usepackage{type1cm}        % activate if the above 3 fonts are
\usepackage{makeidx}         % allows index generation
\usepackage{graphicx}        % standard LaTeX graphics tool
\usepackage{multicol}        % used for the two-column index
\usepackage[bottom]{footmisc}% places footnotes at page bottom

\usepackage{newtxtext}       % 
\usepackage{newtxmath}       % selects Times Roman as basic font

\usepackage{gensymb}
\usepackage[utf8]{inputenc}

\usepackage{hyperref}
\hypersetup{unicode=true,pdfusetitle,
 bookmarks=true,bookmarksnumbered=true,bookmarksopen=true,
 breaklinks=false,pdfborder={0 0 0},backref=false,colorlinks=true,
 linkcolor=blue, citecolor=blue, urlcolor=blue}

\makeindex             % used for the subject index
\begin{document}

\title*{Inducing enantiosensitive permanent multipoles in isotropic samples
with two-color fields}
\author{Andres F. Ordonez and Olga Smirnova}
\institute{Andres F. Ordonez \at Max-Born-Institut, Max-Born-Str. 2A, 12489 Berlin, Germany, \email{ordonez@mbi-berlin.de}
  \and Olga Smirnova \at Max-Born-Institut, Max-Born-Str. 2A, 12489 Berlin, Germany\\
  Technische Universit\"at Berlin, Ernst-Ruska-Geb\"aude, Hardenbergstr. 36A, 10623 Berlin, Germany, \email{smirnova@mbi-berlin.de}}
\maketitle

%% \abstract*{Each chapter should be preceded by an abstract (no more than 200 words) that summarizes the content. The abstract will appear \textit{online} at \url{www.SpringerLink.com} and be available with unrestricted access. This allows unregistered users to read the abstract as a teaser for the complete chapter.
%% Please use the 'starred' version of the \texttt{abstract} command for typesetting the text of the online abstracts (cf. source file of this chapter template \texttt{abstract}) and include them with the source files of your manuscript. Use the plain \texttt{abstract} command if the abstract is also to appear in the printed version of the book.}

\abstract{We find that two-color fields can induce field-free
  permanent dipoles in initially isotropic samples of chiral molecules 
  via resonant electronic excitation in a one-$3\omega$-photon vs.
  three-$\omega$-photons scheme. These permanent dipoles are
  enantiosensitive and can be controlled via the relative phase
  between the two colors.  When the two colors are linearly polarized
  perpendicular to each other, the interference between the two
  pathways induces excitation sensitive to the molecular handedness and  orientation, leading to  uniaxial orientation of the excited molecules and to an enantio-sensitive permanent
  dipole perpendicular to the polarization plane. We also find that
  although a corresponding one-$2\omega$-photon
  vs. two-$\omega$-photons scheme cannot produce enantiosensitive
  permanent dipoles, it can produce enantiosensitive permanent
  quadrupoles that are also controllable through the two-color
  relative phase.}

\section{Introduction }

Chirality (handedness) is the geometrical property that allows us
to distinguish a left hand from a right hand. Like hands, many molecules
have two possible versions which are non-superimposable
mirror images of each other (opposite enantiomers). This ``extra degree of freedom'' stemming
from the reduced symmetry (lack of improper symmetry axes) of chiral
molecules leads to interesting behavior absent in achiral molecules
\cite{condon_theories_1937,soai_asymmetric_1995,fischer_nonlinear_2005,beaulieu_universality_2016,banerjee-ghosh_separation_2018,sanchez_topological_2019}
with profound implications for biology \cite{mason_biomolecular_1988,blackmond_asymmetric_2004}.
Furthermore, since opposite enantiomers share fundamental properties
like their mass and their energy spectrum, one must often rely precisely
on this chiral behavior to tell opposite enantiomers apart
-- a task of immense practical importance in chemistry \cite{lin_chiral_2011,berova_comprehensive_2012}. 

\begin{figure}[b]
\begin{centering}
\includegraphics[scale=0.5]{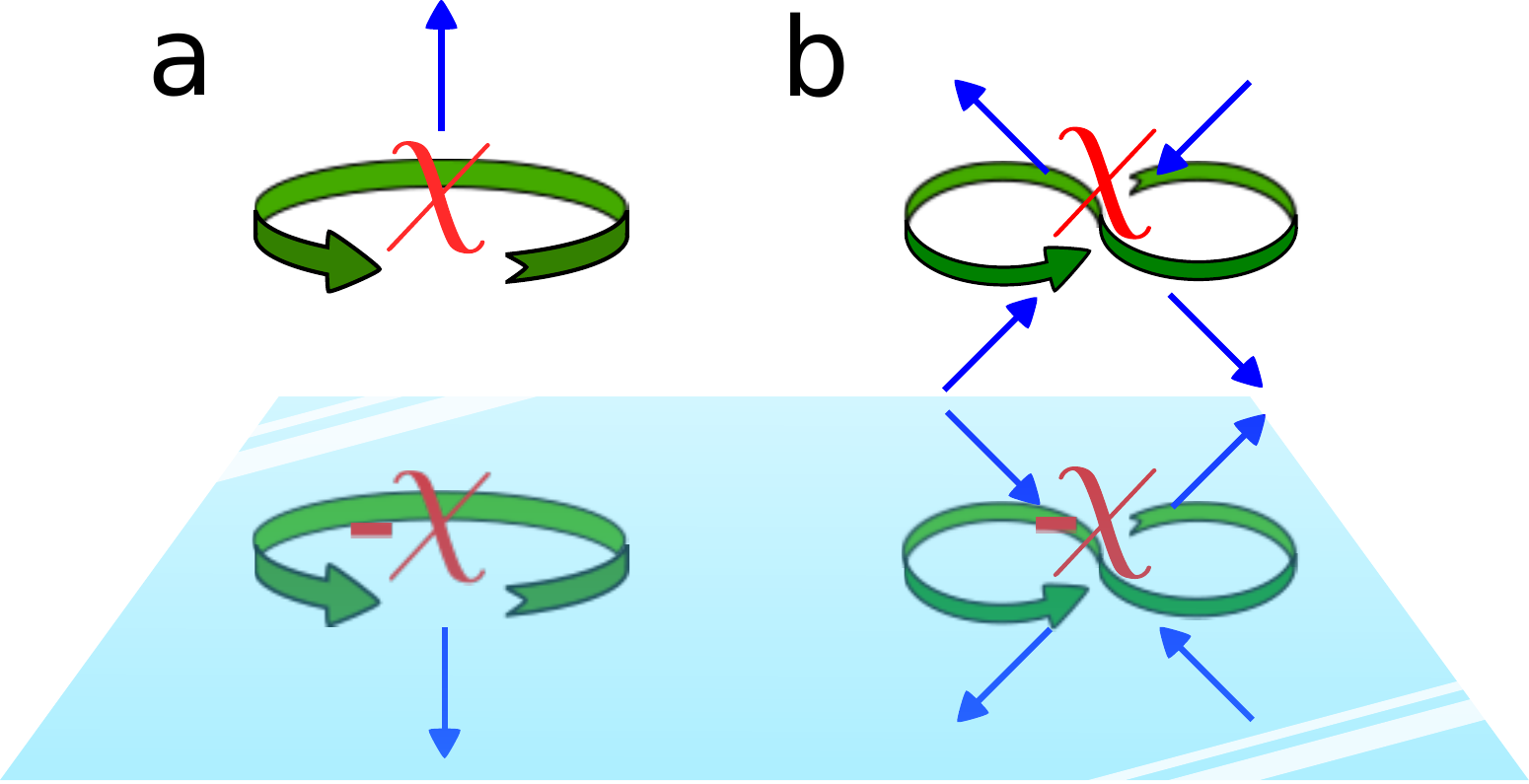}
\par\end{centering}
\caption{Symmetry in $\omega$ and $\omega$-$2\omega$ setups.\textbf{ a}.
A circularly polarized field (circular arrow) interacts with an isotropic
sample of chiral molecules (represented by $\chi$) and produces a
net photoelectron current (in general a vectorial signal) perpendicular
to the polarization plane (arrow pointing up). The mirror reflection
shows that the interaction of the same field with the opposite enantiomer
(represented by $-\chi$) yields the opposite current. \textbf{b.
}A field with its fundamental and second harmonic linearly polarized
perpendicular to each other ($\infty$-like arrow) interacting with
an isotropic chiral sample produces a quadrupolar photoelectron current
(in general a quadrupolar signal). The mirror reflection shows that
the interaction of the same field with the opposite enantiomer yields
the opposite current. In both \textbf{a} and \textbf{b}, the reversal of
the signal when the polarization is changed follows from considering a
rotation of $180^{\circ}$ (not shown) of the full system, which changes the
polarization but not the isotropic sample (see Figs. 2-4 in Ref.
\cite{ordonez_generalized_2018-1}). }
\label{fig:PECD}
\end{figure}

An example of this chiral behavior is the phenomenon known as
photoelectron circular dichroism (PECD)
\cite{ritchie_theory_1976,bowering_asymmetry_2001,powis_photoelectron_2008,beaulieu_universality_2016},
%which manifests in the generation of a net 
which consists in the generation of a net
photoelectron current from
an isotropic sample of chiral molecules irradiated by circularly polarized
light \cite{ordonez_generalized_2018-1, ordonez_propensity_2019-1, ordonez_propensity_2019}. This photoelectron current,
which results from different amounts of photoelectrons being emitted
in opposite directions, is directed along the normal to the
polarization plane (because of the overall cylindrical symmetry) and
changes sign when either the enantiomer or the circular polarization
is reversed (see Fig. \ref{fig:PECD}a). Importantly, PECD occurs
within the electric-dipole approximation, which makes typical PECD
signals orders of magnitude stronger than traditional enantiosensitive signals,
such as circular dichroism (CD), which rely on interactions beyond the
electric-dipole approximation
\cite{barron_optical_1979,berova_comprehensive_2012}.  Furthermore,
the electric-dipole approximation also rules out any influence of the
wave vector of the incident light and hence of the momentum of the
photons.

Given that: the molecules are randomly oriented in space, the
electric field is circularly polarized, and the momentum of the
photon does not play any role in PECD; it is only natural to wonder
\emph{why does a net current of photoelectrons perpendicular to the
  polarization plane occur?} From the point of view of symmetry, the
question would be instead \emph{what symmetry prevents this current
  from taking place in the case of achiral molecules?} The answer is
simple: in the electric-dipole approximation\footnote{Beyond the
  electric-dipole approximation the wave vector of the light breaks
  reflection symmetry.}  the system consisting of isotropic achiral
molecules together with the circularly polarized electric field is
symmetric with respect to reflection in the polarization
plane\footnote{Note that circularly polarized light is not chiral
  within the electric-dipole approximation, and therefore the
  chirality of the light itself does not play a role in PECD
  \cite{ordonez_generalized_2018-1}.} and therefore the current normal to the polarization
plane must vanish.  When achiral molecules are replaced by chiral
molecules, this mirror symmetry is broken and the PECD current emerges
\cite{ordonez_generalized_2018-1}.

While this symmetry analysis does not provide an 
answer in terms of the specific
mechanism, the insight it provides applies to several other closely
related effects occurring within the electric dipole approximation,
which rely on electric field polarizations confined to a plane and
yield enantiosensitive vectorial responses perpendicular to that plane
\cite{giordmaine_nonlinear_1965,rentzepis_coherent_1966,fischer_nonlinear_2005,patterson_enantiomer-specific_2013,patterson_sensitive_2013,yachmenev_detecting_2016,beaulieu_photoexcitation_2018,ordonez_generalized_2018-1}.
For example, if the photon energy of the circularly polarized light is
not enough to ionize the molecule, the lack of reflection symmetry due
to the chiral molecules leads to oscillating bound currents normal to
the polarization plane
\cite{beaulieu_photoexcitation_2018,ordonez_generalized_2018-1}.  In
this case, the current results from the excitation of bound states and
the associated oscillation of the expected value of the electric
dipole operator. The enantiosensitivity is reflected in the phase
of the oscillations, which are out of phase in opposite enantiomers.

Analogously, one may also expect that it should be 
possible to induce
permanent electric dipoles (i.e. non-vanishing zero-frequency
components of the expected value of the electric dipole operator)
normal to the polarization plane and with opposite directions for
opposite enantiomers. Indeed, such static electric dipoles have been investigated in the context of optical rectification \cite{koroteev_new_1994, wozniak_non-resonant_1995, zawodny_optical_1996, wozniak_optical_1997, fischer_optical_2002}, where two excited states close in energy are resonantly excited with monochromatic circularly polarized light. Very recently enantiosensitive static dipoles have also been studied in the context of molecular orientation induced by
intense off-resonant light pulses \cite{gershnabel_orienting_2018,tutunnikov_selective_2018,tutunnikov_laser-induced_2019,milner_controlled_2019}. Such light pulses excite rotational dynamics and
%, unlike the optical rectification scheme, 
cause orientation of one of the
molecular axes that persists after the pulse is over.
Here we show that field-free 
enantiosensitive permanent electric dipoles and
the associated orientation can also be induced in the context of
purely electronic excitation on ultrafast time-scales, without  
relying on rotational dynamics. %This 
%enantio-sensitive excitation is 
We achieve this via 
interference of one- and three-photon excitation pathways.

Quite recently an extension of single-color PECD to
two-color $\omega$-$2\omega$ fields with orthogonal linear polarizations has been observed 
\cite{demekhin_photoelectron_2018, demekhin_photoelectron_2019, rozen_controlling_2019} (see Fig. \ref{fig:PECD}b).
As we discuss in Ref. \cite{ordonez_tensorial}, this is an example of
how molecular chirality can be reflected not only in scalar (e.g.
CD) and vectorial observables (e.g. PECD), 
but also in higher-rank tensor observables. 
Here we show that two-color $\omega$-$2\omega$
fields with linear polarizations perpendicular to each other can
induce enantiosensitive permanent quadrupoles in samples of
isotropic chiral molecules. 
  %%, analogous to the quadrupolar currents
  %% found in Refs. \cite{demekhin_photoelectron_2018,
  %%   demekhin_photoelectron_2019, rozen_controlling_2019}

\section{Exciting enantiosensitive permanent dipole\label{subsec:Permanent-dipole}}

\begin{figure}[b]
\sidecaption
%% \begin{centering}
\includegraphics[width=0.09\paperwidth]{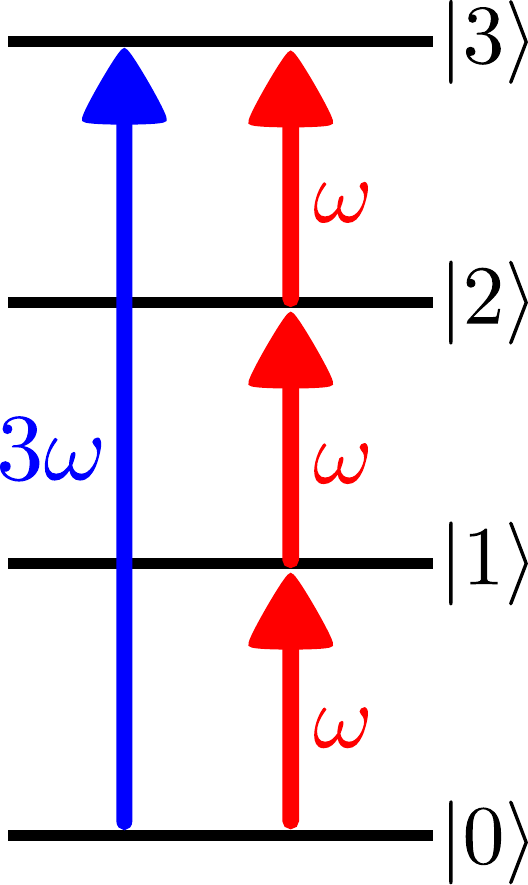}
%% \par\end{centering}
\caption{Excitation scheme used to produce an enantiosensitive permanent dipole
in an isotropic sample of chiral molecules.\label{fig:scheme_dipole}}
\end{figure}

Consider the excitation scheme depicted in Fig. \ref{fig:scheme_dipole},
where the interference of contributions from a one-$3\omega$-photon
pathway and a three-$\omega$-photon pathway control the population
of the state $\vert 3 \rangle$ of a chiral molecule. For simplicity, we first consider excitation via intermediate resonances in states $\vert 1 \rangle$ and $\vert 2 \rangle$. The presence of resonances in these states is not essential, as discussed later, but simplifies the analysis.  The field is assumed
to have the form 
\begin{equation}
\vec{E}^{\mathrm{L}}\left(t\right)=F\left(t\right)\left(\vec{E}_{\omega}^{\mathrm{L}}e^{-i\omega t}+\vec{E}_{3\omega}^{\mathrm{L}}e^{-3i\omega t}\right)+\mathrm{c.c.},
\end{equation}
where $F\left(t\right)$ is a smooth envelope,
$\vec{E}_{\omega}^{\mathrm{L}}$ and $\vec{E}_{3\omega}^{\mathrm{L}}$
specify the polarizations and phases of each frequency, and the
$\mathrm{L}$ and $\mathrm{M}$ superscripts indicate vectors and
functions in the laboratory frame and in the molecular frame,
respectively. For a given molecular orientation
$\varrho\equiv\alpha\beta\gamma$, where $\alpha\beta\gamma$ are the
Euler angles, the wave function after the interaction is
\begin{equation}
\Psi^{\mathrm{M}}(\vec{r}^{\mathrm{M}},\varrho)=\sum_{i=0}^{3}a_{i}\left(\varrho\right)e^{-i\omega_{i}t}\psi_{i}^{\mathrm{M}}(\vec{r}^{\mathrm{M}}),\label{eq:wavefunction}
\end{equation}
where $\psi_{i}^{\mathrm{M}}(\vec{r}^{\mathrm{M}})$ is the coordinate representation of state $\vert i\rangle$ in the molecular frame. In the perturbative regime we have %take $a_{0}=1$ and 
\begin{equation}
a_{3}\left(\varrho\right)=A_{3}^{\left(1\right)}[\vec{d}_{3,0}^{\mathrm{L}}\left(\varrho\right)\cdot\vec{E}_{3\omega}^{\mathrm{L}}]+A_{3}^{\left(3\right)}[\vec{d}_{3,2}^{\mathrm{L}}\left(\varrho\right)\cdot\vec{E}_{\omega}^{\mathrm{L}}][\vec{d}_{2,1}^{\mathrm{L}}\left(\varrho\right)\cdot\vec{E}_{\omega}^{\mathrm{L}}][\vec{d}_{1,0}^{\mathrm{L}}\left(\varrho\right)\cdot\vec{E}_{\omega}^{\mathrm{L}}],
\label{eq:a3}
\end{equation}
where $A_{3}^{\left(1\right)}$ and $A_{3}^{\left(3\right)}$ are first- and
third-order coupling constants that depend on the detunings and the envelope (see Appendix). Analogous expressions apply for the other amplitudes $a_i$. 
The transition
dipoles 
$\vec{d}_{i,j}^{\mathrm{M}}\equiv\langle\psi_{i}^{\mathrm{M}}(\vec{r}^{\mathrm{M}})\vert\vec{d}^{\mathrm{M}}\vert\psi_{j}^{\mathrm{M}}(\vec{r}^{\mathrm{M}})\rangle$
are fixed in the molecular frame and have been expressed in the laboratory
frame using the rotation matrix $R\left(\varrho\right)$ according
to $\vec{d}_{i,j}^{\mathrm{L}}\left(\varrho\right)=R\left(\varrho\right)\vec{d}_{i,j}^{\mathrm{M}}$. 

The expected value of the electric dipole operator in the molecular
frame $\langle\vec{d}^{\mathrm{M}}\left(\varrho\right)\rangle$ $\equiv$
$\langle\Psi^{\mathrm{M}}(\vec{r}^{\mathrm{M}},\varrho)\vert$ $\vec{d}^{\mathrm{M}}$
$\vert\Psi^{\mathrm{M}}\left(\vec{r}^{\mathrm{M}},\varrho\right)\rangle$
has a zero-frequency component of the form
\begin{equation}
\langle\vec{d}^{\mathrm{M}}\left(\varrho\right)\rangle_{\omega=0} =\sum_{i=0}^{3}\left|a_{i}\left(\varrho\right)\right|^{2}\vec{d}_{i,i}^{\mathrm{M}}
%=\left|a_{3}\left(\varrho\right)\right|^{2}\vec{d}_{3,3}^{\mathrm{M}}
,
\label{eq:d_m}
\end{equation}
%i.e. the permanent dipole is excited in each of the states populated after the end of the pulse:
i.e. the permanent dipole for a given molecular orientation is the sum of the permanent dipoles of each state weighted by their orientation-dependent populations at the end of the pulse. 
% \begin{equation}
% \langle\vec{d}_1^{\mathrm{M}}\left(\varrho\right)\rangle_{\omega=0} 
% =\left|a_{1}\left(\varrho\right)\right|^{2}\vec{d}_{1,1}^{\mathrm{M}},\label{eq:d_m1}
% \end{equation}
% \begin{equation}
% \langle\vec{d}_2^{\mathrm{M}}\left(\varrho\right)\rangle_{\omega=0} 
% =\left|a_{2}\left(\varrho\right)\right|^{2}\vec{d}_{2,2}^{\mathrm{M}},\label{eq:d_m2}
% \end{equation}
% \begin{equation}
% \langle\vec{d}_3^{\mathrm{M}}\left(\varrho\right)\rangle_{\omega=0} 
% =\left|a_{3}\left(\varrho\right)\right|^{2}\vec{d}_{3,3}^{\mathrm{M}},\label{eq:d_m3}
% \end{equation}
% \begin{equation}
% \langle\vec{d}_0^{\mathrm{M}}\left(\varrho\right)\rangle_{\omega=0} =
% \left(1-\left|a_{1}\left(\varrho\right)\right|^{2}-\left|a_{2}\left(\varrho\right)\right|^{2}-\left|a_{3}\left(\varrho\right)\right|^{2}\right)\vec{d}_{0,0}^{\mathrm{M}},
% \label{eq:d_m0}
% \end{equation}
% The total dipole in Eq. (\ref{eq:d_m}) is non-zero, since the permanent dipoles $\vec{d}_{i,i}^{\mathrm{M}}$ of the different electronic states $\vert0\rangle$, $\vert1\rangle$,  $\vert2\rangle$, and $\vert3\rangle$ are, in general, different.

Transforming $\langle\vec{d}^{\mathrm{M}}\left(\varrho\right)\rangle_{\omega=0}$
to the laboratory frame and averaging over all molecular orientations
%\footnote{We use Eq. (A16) in Ref. \cite{ordonez_generalized_2018-1} for the interference
% term; the direct terms vanish by symmetry.} 
yields the permanent dipole

\begin{equation}
    \langle \vec{d}^\mathrm{L} \rangle_{\omega=0} \equiv \int \mathrm{d}\varrho \langle \vec{d}^\mathrm{L} (\varrho) \rangle_{\omega=0}.
    \label{eq:d_total}
\end{equation}

The contribution of state $\vert3\rangle$ to this expression reads as\footnote{We use Eq. (A16) in Ref. \cite{ordonez_generalized_2018-1} for the interference
term. The direct terms vanish because the possible non-zero field pseudovectors are purely imaginary, e.g. $(\vec{E}_{3\omega}^*\times\vec{E}_{3\omega})$, while the accompanying molecular pseudoscalars are real and the coupling coefficients appear within absolute values, see Ref. \cite{andrews_threedimensional_1977}.}
\begin{equation}
\langle\vec{d}_3^{\mathrm{L}}\rangle_{\omega=0} \equiv
\int\mathrm{d}\varrho\,\left|a_{3}\left(\varrho\right)\right|^{2}\vec{d}_{3,3}^{\mathrm{L}}\left(\varrho\right)
=A_3^{\left(1\right)*}A_3^{\left(3\right)}\chi_3\vec{Z}^{\mathrm{L}}+\mathrm{c.c.},\label{eq:d_solved}
\end{equation}
where $\int\mathrm{d}\varrho\equiv\int_{0}^{2\pi}\mathrm{d}\alpha\int_{0}^{\pi}\mathrm{d}\beta\int_{0}^{2\pi}\mathrm{d}\gamma/8\pi^{2}$ is the integral over all molecular orientations 
 and we defined 
\begin{equation}
\chi_i\equiv\frac{1}{30}\left[(\vec{d}_{2,1}^{\mathrm{M}}\cdot\vec{d}_{1,0}^{\mathrm{M}})\vec{d}_{3,2}^{\mathrm{M}}+(\vec{d}_{3,2}^{\mathrm{M}}\cdot\vec{d}_{1,0}^{\mathrm{M}})\vec{d}_{2,1}^{\mathrm{M}}+(\vec{d}_{3,2}^{\mathrm{M}}\cdot\vec{d}_{2,1}^{\mathrm{M}})\vec{d}_{1,0}^{\mathrm{M}}\right]\cdot(\vec{d}_{3,0}^{\mathrm{M}}\times\vec{d}_{i,i}^{\mathrm{M}}),\label{eq:chi_m}
\end{equation}
\begin{equation}
\vec{Z}^{\mathrm{L}}\equiv\left(\vec{E}_{\omega}^{\mathrm{L}}\cdot\vec{E}_{\omega}^{\mathrm{L}}\right)\left(\vec{E}_{\omega}^{\mathrm{L}}\times\vec{E}_{3\omega}^{\mathrm{L}*}\right).\label{eq:Z_l}
\end{equation}
%The third order coupling constant $A^{(3)}$  for the case of resonant intermediate states is $A^{(3)}=i^3\int_{-\infty}^{\infty}dt_3 f(t_3)\int_{-\infty}^{t_3}dt_2 f(t_2)\int_{-\infty}^{t_2}dt_1 f(t_1)$ where $f(t)$ is a pulse envelope. 
$\chi_3$ is a rotationally invariant molecular pseudoscalar, i.e. a molecular quantity independent of the molecular orientation. It has opposite signs for opposite enantiomers and 
vanishes for achiral molecules;
$\chi_3$ encodes the enantiosensitivity of $\langle\vec{d}^{\mathrm{L}}_3\rangle_{\omega=0}$.
Selection rules for $\chi_3$ can be directly read off from Eq. (\ref{eq:chi_m}).
In particular, it vanishes if $\vec{d}_{3,0}^{\mathrm{M}}$ and $\vec{d}_{3,3}^{\mathrm{M}}$
are collinear. $\vec{Z}^{\mathrm{L}}$ 
is a light pseudovector -- it
is a vector that depends only on the light's polarization and
is invariant under the inversion operation; $\vec{Z}^{\mathrm{L}}$
determines the direction of $\langle\vec{d}^{\mathrm{L}}_3\rangle_{\omega=0}$. Selection rules for 
$\vec{Z}^{\mathrm{L}}$ can be read off directly from Eq. (\ref{eq:Z_l}). In particular, it vanishes if $\omega$ is circularly polarized
($\vec{E}_{\omega}^{\mathrm{L}}\cdot\vec{E}_{\omega}^{\mathrm{L}}=0$)
or if $\omega$ and $3\omega$ are linearly polarized parallel
to each other ($\vec{E}_{\omega}^{\mathrm{L}}\times\vec{E}_{3\omega}^{\mathrm{L}*}=0$). 

For example, if we choose $\omega$ and $3\omega$ linearly polarized
perpendicular to each other, say
$\vec{E}_{\omega}^{\mathrm{L}}=\vec{x}^{\mathrm{L}}$ and
$\vec{E}_{3\omega}^{\mathrm{L}}=e^{-i\phi}\vec{y}^{\mathrm{L}}$, with
$\vec{x}^{\mathrm{L}}$ and $\vec{y}^{\mathrm{L}}$ the unitary vectors along each axis then
\begin{equation}
\vec{E}^{\mathrm{L}}\left(t\right)=2F\left(t\right)\left[\cos\left(\omega t\right)\vec{x}^{\mathrm{L}}+\cos\left(3\omega t+\phi\right)\vec{y}^{\mathrm{L}}\right]\label{eq:E_xy}
\end{equation}
and we obtain
%the expected value of the electric dipole induced in the initially
%isotropic sample yields
\begin{equation}
\langle\vec{d}_{3}^{\mathrm{L}}\rangle_{\omega=0}=2\chi_3\Re\left\{ A_3^{\left(1\right)*}A_3^{\left(3\right)}e^{i\phi}\right\} \vec{z}^{\mathrm{L}},
\label{eq:d_3_solved}
\end{equation}
% \begin{equation}
% \langle\vec{d}_{3}^{\mathrm{L}}\rangle_{\omega=0}= A_3^{\left(1\right)*}A_3^{\left(3\right)} e^{i\phi} \chi_3 \vec{z}^{\mathrm{L}} + \mathrm{c.c.},
% \label{eq:d_3_solved}
% \end{equation}
% \begin{equation}
% \langle\vec{d}^{\mathrm{L}}_3\rangle_{\omega=0}=2\chi_3\cos\phi \vec{z}^{\mathrm{L}},
% \end{equation}
i.e., $\langle\vec{d}^{\mathrm{L}}_3\rangle_{\omega=0}$ is perpendicular
to the polarization plane and its magnitude and sign can be controlled
through the relative phase $\phi$. Note that the relative phase of the coupling coefficients $A_3^{(1)}$ and $A_3^{(3)}$, which can be modified for example by changing the detunings, must also be taken into account. 

The contributions from states $\vert1\rangle$, and $\vert2\rangle$ to the permanent dipole (\ref{eq:d_total}) have the same structure as Eq. (\ref{eq:d_3_solved}), albeit with different coupling constants and molecular pseudoscalars $\chi_1$ and $\chi_2$, respectively [see Eq. (\ref{eq:chi_m})]. Since $\vert a_0\vert^2 = 1 - \vert a_1 \vert^2 - \vert a_2 \vert^2 - \vert a_3\vert^2$, the contribution from the ground state involves the coupling constants associated to $\vert1\rangle$, $\vert2\rangle$, and $\vert 3 \rangle$, and a molecular pseudoscalar $\chi_0$ [see Eq. (\ref{eq:chi_m})]. Together, these contributions yield
% For the particular polarization in Eq. (\ref{eq:E_xy}), all the light pseudovectors yield $e^{i\phi}\vec{z}^{\mathrm{L}}$ and the molecular pseudoscalars are given by Eq. (\ref{eq:chi_m}) for $i=0,1,2,3$, so that the total dipole in Eq. (\ref{eq:d_total}) yields 
% \begin{align}
%     \langle \vec{d}^\mathrm{L} \rangle_{\omega=0} =& \bigg[A_{1}^{\left(1\right)}A_{1}^{\left(3\right)*}\left(\chi_{1}-\chi_{0}\right)+A_{2}^{\left(2\right)\prime*}A_{2}^{\left(2\right)}\left(\chi_{2}-\chi_{0}\right)\nonumber\\
%     &+A_{3}^{\left(1\right)*}A_{3}^{\left(3\right)}\left(\chi_{3}-\chi_{0}\right)\bigg]e^{i\phi}\vec{z}^{\mathrm{L}}+\mathrm{c.c.}
%     \label{eq:d_total_solved}
% \end{align}
\begin{align}
    \langle \vec{d}^\mathrm{L} \rangle_{\omega=0} =& 2\bigg[(\chi_{1}-\chi_{0}) \Re\left\{A_{1}^{(1)} A_{1}^{(3)*}e^{i\phi}\right\}
    +(\chi_{2}-\chi_{0})\Re\left\{A_{2}^{(2)\prime*} A_{2}^{(2)}e^{i\phi}\right\}\nonumber\\
    &+(\chi_{3}-\chi_{0})\Re\left\{A_{3}^{(1)*}  A_{3}^{(3)}e^{i\phi}\right\}\bigg]\vec{z}^{\mathrm{L}}
    \label{eq:d_total_solved}
\end{align}
where $A_1^{(1)}$ and $A_{1}^{(3)}$ are the coupling coefficients for the transitions $\vert 0 \rangle$ $\stackrel{\omega}{\rightarrow}$ $\vert 1 \rangle$ 
and $\vert 0\rangle$ $\stackrel{3\omega}{\rightarrow}$ $\vert 3 \rangle$ $\stackrel{-\omega}{\rightarrow}$ $\vert 2 \rangle$ $\stackrel{-\omega}{\rightarrow}$ $\vert 1 \rangle$, respectively; 
 $A_2^{(2)}$ and $A_{2}^{(2)\prime}$ are the coupling coefficients for the transitions $\vert 0 \rangle$ $\stackrel{\omega}{\rightarrow}$ $\vert 1 \rangle$ $\stackrel{\omega}{\rightarrow}$ $\vert 2 \rangle$ 
and $\vert 0 \rangle$ $\stackrel{3\omega}{\rightarrow}$ $\vert 3 \rangle$ $\stackrel{-\omega}{\rightarrow}$ $\vert 2 \rangle$, respectively.
%; and we took into account that $\vert a_0\vert^2 = 1 - \vert a_1 \vert^2 - \vert a_2 \vert^2 - \vert a_3\vert^2$. 

In the absence of the intermediate resonances through the states $|1\rangle$ and $|2\rangle$ the contribution from the third-order term in Eq. (\ref{eq:a3}) turns into a sum over all intermediate states $|j\rangle$ and $|k\rangle$ weighted by a coefficient $A^{(3)}_{3;jk}$. 
The intermediate states retain no population at the end of the pulse and the permanent dipole takes the form 
\begin{align}
    \langle \vec{d}^\mathrm{L} \rangle_{\omega=0} &= 2\sum_{j,k} (\chi_{3;jk}-\chi_{0;jk}) \Re\left\{A_{3}^{(1)*}A_{3;jk}^{(3)} \vec{Z}^{\mathrm{L}}\right\}\nonumber\\
    &= 2F_{0}^{4}\left[2\pi \delta_{\sigma}(\Delta)\right]^{2}\sum_{j,k}\frac{\chi_{3;jk}-\chi_{0;jk}}{\left(\omega_{k,0}-2\omega_{L}\right)\left(\omega_{j,0}-\omega_{L}\right)}\Re\left\{ \vec{Z}^{\mathrm{L}}\right\} 
    \label{eq:d_total_solved_offres}
\end{align}
which is valid for arbitrary polarizations [see Eq. (\ref{eq:Z_l})]. Here $\chi_{i;jk}$ is given by Eq. (\ref{eq:chi_m}) with the replacements $1\rightarrow j$ and $2\rightarrow k$. In the second equality we wrote the coupling constants explicitly, $\omega_{i,j}\equiv\omega_i-\omega_j$, $\Delta\equiv\omega_{3,0}-3\omega_{L}$, and we took $\int_{-\infty}^{\infty}\mathrm{d}t\,F(t)e^{i\omega t}\equiv 2\pi F_0 \delta_{\sigma}(\omega)$ with $\delta_{\sigma}(\omega)$ equal to the Dirac delta in the limit of infinitesimal $\sigma$.

\subsubsection*{A simple picture of the mechanism leading to the enantiosensitive
permanent dipole}

\begin{figure}[b]
\begin{centering}
\includegraphics[scale=0.25]{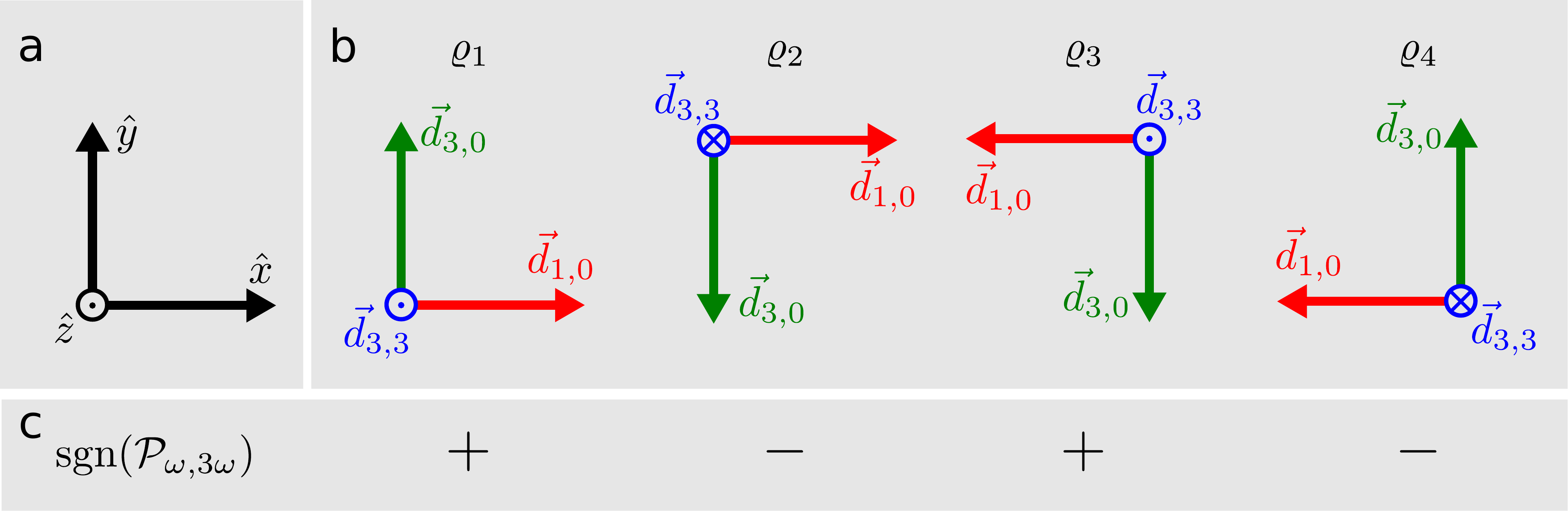}
\par\end{centering}
\caption{Simple analysis of the mechanism leading to an enantiosensitive permanent
dipole for a field (\ref{eq:E_xy}) and a dummy molecule with $\vec{d}_{1,0}$
and $\vec{d}_{3,0}$, perpendicular to each other and $\vec{d}_{3,2}=\vec{d}_{2,1}=\vec{d}_{1,0}$.
Only the component of $\vec{d}_{3,3}$ perpendicular to the plane
defined by $\vec{d}_{1,0}$ and $\vec{d}_{3,0}$ is shown. \textbf{a.
}Laboratory frame. \textbf{b}. Molecular orientations with $\vec{d}_{1,0}$
aligned along $\vec{x}$ and $\vec{d}_{3,0}$ aligned along $\vec{y}$.
 \textbf{c. }Sign of the interference term (\ref{eq:Pw3w}) for each
molecular orientation. The interference distinguishes orientations
$\varrho_{1}$ and $\varrho_{3}$ from orientations $\varrho_{2}$
and $\varrho_{4}$ and therefore causes the molecular axis $\vec{d}_{3,3}$
to become oriented. This leads to a non-vanishing permanent dipole.\label{fig:Simple-picture}}
\end{figure}

The orientation averaging procedure we applied
\cite{andrews_threedimensional_1977}, although very powerful, is also
rather formal. Below we demonstrate that the mechanism leading to the generation  of the permanent dipole $\langle\vec{d}^{\mathrm{L}}\rangle_{\omega=0}$ 
% is the  sensitivity of excitation to the molecular orientation and handedness, resulting in the oriented ensemble of excited molecules, with their orientations opposite for the opposite enantiomers. 
stems from the sensitivity of the 
excitation to the molecular orientation and handedness, which induces uniaxial and enantiosensitive orientation of the initially isotropic sample. We remark that the excitation induces \emph{orientation} ($\uparrow$) as opposed to just \emph{alignment} ($\updownarrow$) and that this orientation is furthermore enantiosensitive. 
% This is very different from the well-known alignment-inducing excitation: here we induce enantio-sensitive orientation.
%To shed some light on the mechanism leading to the
%permanent dipole %$\langle\vec{d}^{\mathrm{L}}\rangle_{\omega=0}$,

Consider the interaction of the field (\ref{eq:E_xy}) with a dummy
molecule with $\vec{d}_{1,0}^{\mathrm{M}}$ and
$\vec{d}_{3,0}^{\mathrm{M}}$ perpendicular to each other and
$\vec{d}_{3,2}^{\mathrm{M}}=\vec{d}_{2,1}^{\mathrm{M}}=\vec{d}_{1,0}^{\mathrm{M}}$. For simplicity we again assume that intermediate states are resonantly excited and that only state $|3\rangle$ has a non-zero permanent dipole. 
% In the general case of non-resonant excitation of intermediate states, the sum over such states and their respective detunings will contribute to the expressions below, as shown in Eq. (\ref{eq:d_total_solved_offres}).
%% For such molecule
%% $\chi=\frac{1}{10}\vert\vec{d}_{1,0}^{\mathrm{M}}\vert^{2}[\vec{d}_{3,3}^{\mathrm{M}}\cdot(\vec{d}_{1,0}^{\mathrm{M}}\times\vec{d}_{3,0}^{\mathrm{M}})]$.
The population $P_{3}\left(\varrho\right)\equiv\vert
a_{3}\left(\varrho\right)\vert^{2}$ of the excited state $\vert 3
\rangle$ reads  {[}see Eq. (\ref{eq:a3}){]}
% \begin{equation}
% P_{3}\left(\varrho\right)=-F^{2}\left(t\right)\left[ \mathcal{P}_{\omega}\left(\varrho\right)+ \mathcal{P}_{3\omega}\left(\varrho\right)+2\Re\left \{ e^{i\phi}\right\} \mathcal{P}_{\omega,3\omega}\left(\varrho\right)\right] 
% \end{equation}
\begin{equation}
P_{3}\left(\varrho\right)=\vert A_3^{\left(1\right)}\vert^{2}\mathcal{P}_{3\omega}\left(\varrho\right)+\vert A_3^{\left(3\right)}\vert^{2}\mathcal{P}_{\omega}\left(\varrho\right)+2\Re\left\{ A_3^{\left(1\right)*}A_3^{\left(3\right)}e^{i\phi}\right\} \mathcal{P}_{\omega,3\omega}\left(\varrho\right) 
\end{equation}
where 
\begin{equation}
\mathcal{P}_{\omega}\left(\varrho\right) \equiv\left[\vec{d}_{1,0}^{\mathrm{L}}\left(\varrho\right)\cdot\vec{x}^{\mathrm{L}}\right]^{6}, \quad \mathcal{P}_{3\omega}\left(\varrho\right)\equiv\left[\vec{d}_{3,0}^{\mathrm{L}}\left(\varrho\right)\cdot\vec{y}^{\mathrm{L}}\right]^{2},
\end{equation}
\begin{equation}
\mathcal{P}_{\omega,3\omega}\left(\varrho\right)\equiv\left[\vec{d}_{3,0}^{\mathrm{L}}\left(\varrho\right)\cdot\vec{y}^{\mathrm{L}}\right]\left[\vec{d}_{1,0}^{\mathrm{L}}\left(\varrho\right)\cdot\vec{x}^{\mathrm{L}}\right]^{3}.\label{eq:Pw3w}
\end{equation}
$\mathcal{P}_{\omega}$ will select molecular
orientations where $\vec{d}_{1,0}^{\mathrm{L}}$ is aligned along the
$\vec{x}^{\mathrm{L}}$
axis. $\mathcal{P}_{3\omega}$ will select
molecular orientations where $\vec{d}_{3,0}^{\mathrm{L}}$ is aligned
along the $\vec{y}^{\mathrm{L}}$
axis. $\mathcal{P}_{\omega,3\omega}$ will
select molecular orientations where $\vec{d}_{1,0}^{\mathrm{L}}$
is aligned along the $\vec{x}^{\mathrm{L}}$ axis \emph{and}
$\vec{d}_{3,0}^{\mathrm{L}}$ is aligned along the
$\vec{y}^{\mathrm{L}}$ axis. These orientations are shown in
Fig. \ref{fig:Simple-picture}b. While the direct terms
$\mathcal{P}_{\omega}$ and $\mathcal{P}_{3\omega}$ do not distinguish
between this subset of orientations $\{\varrho_i\}_{i=1}^{4}$, the
interference term $\mathcal{P}_{\omega,3\omega}$ will be positive for
orientations $\varrho_{1}$ and $\varrho_{3}$ and negative for
orientations $\varrho_{2}$ and $\varrho_{4}$. This produces an
imbalance between the number of excited molecules with orientations
$\varrho_{1}$ and $\varrho_{3}$ and those with orientations
$\varrho_{2}$ and $\varrho_{4}$. As can be seen in
Fig. \ref{fig:Simple-picture}, this imbalance amounts to the molecular
axis $\vec{d}_{1,0}^{\mathrm{M}}\times\vec{d}_{3,0}^{\mathrm{M}}$
being oriented. That is, the field (\ref{eq:E_xy}) induces field-free
uniaxial orientation of the molecular sample in the state $|3\rangle$. The emergence of a permanent
dipole follows trivially, provided that 
$\vec{d}_{3,3}^{\mathrm{M}}$ has a
non-zero component along the oriented axis, i.e. as long as
$\vec{d}_{3,3}^{\mathrm{M}}\cdot(\vec{d}_{1,0}^{\mathrm{M}}\times\vec{d}_{3,0}^{\mathrm{M}})\neq0$.
Note that, according to Eq. (\ref{eq:chi_m}),  this is in agreement with having $\chi_3\neq 0$.
If we consider the situation depicted in Fig. \ref{fig:Simple-picture}
now mirror reflected across the polarization plane, which is
equivalent to swapping the enantiomer while leaving the field as it
is, we immediately see that $\vec{d}_{3,3}^{\mathrm{M}}$ and therefore
also $\langle\vec{d}_3^{\mathrm{L}}\rangle_{\omega=0}$ point in the
opposite direction, which explains the enantiosensitivity of
$\langle\vec{d}_3^{\mathrm{L}}\rangle_{\omega=0}$.

Since the emergence of a permanent dipole $\langle\vec{d}^{\mathrm{L}}\rangle_{\omega=0}$
relies on the molecules in the excited state $\vert 3 \rangle$ being oriented,
we expect $\langle\vec{d}^{\mathrm{L}}\rangle_{\omega=0}$ to survive
for at least a few picoseconds before decaying due to molecular rotation.
A decay of the dipole on the picosecond time-scale should lead to broadband THz
emission \cite{cook_intense_2000} with an enantiosensitive phase.
Furthermore, a quantum treatment of the rotational dynamics might
reveal revivals of the molecular orientation (see e.g. Ref. \cite{tutunnikov_laser-induced_2019}) . 

% Admittedly, strong orientation by excitation requires large excitation probabilities, which are not possible in the perturbative regime of excitation. However, large excitation probabilities are readily achievable in strong laser fields. 
% Moreover, in such fields resonances are ubiquitous due to the ponderomotive shift of excited states. The resonances become inevitable as soon as the ponderometive potential 
% $E^2/(4\omega^2)$ becomes comparable to the 
% laser frequency $\omega$.

\section{Exciting enantiosensitive permanent quadrupole}\label{sec:quadrupole}

\begin{figure}[b]
\sidecaption
%% \begin{centering}
\includegraphics[width=0.09\paperwidth]{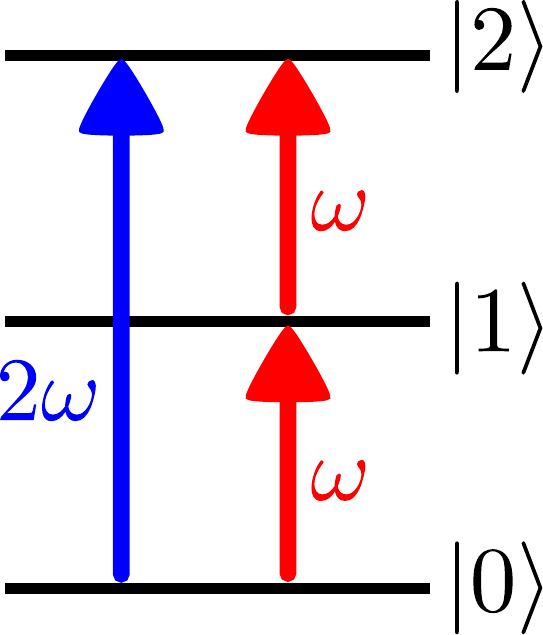}
%% \par\end{centering}
\caption{Excitation scheme used to produce an enantiosensitive permanent quadrupole
in an isotropic sample of chiral molecules.\label{fig:scheme_quadrupole}}
\end{figure}

Let us now consider the control scheme depicted in Fig. \ref{fig:scheme_dipole},
where the interference of contributions from a one-$2\omega$-photon
pathway and a two-$\omega$-photon pathway control the population
of the state $\vert2\rangle$ of a chiral molecule. In this case the field
reads as
\begin{equation}
\vec{E}^{\mathrm{L}}\left(t\right)=F\left(t\right)\left(\vec{E}_{\omega}^{\mathrm{L}}e^{-i\omega t}+\vec{E}_{2\omega}^{\mathrm{L}}e^{-2i\omega t}\right)+\mathrm{c.c.}
\end{equation}
As in the previous section we begin assuming an intermediate resonance and then consider the case where the intermediate state is not resonant.
% For simplicity we begin consider resonant populations of intermediate states. In a general case of virtual intermediate states, the sum over all possible  states with the corresponding detunings should be included, as discussed in the Outlook section.  
The wave function reads as in Eq. (\ref{eq:wavefunction}) but
with a sum up to $i=2$,
\begin{equation}
a_{2}\left(\varrho\right) = A_2^{\left(1\right)}[\vec{d}_{2,0}^{\mathrm{L}}\left(\varrho\right)\cdot\vec{E}_{2\omega}^{\mathrm{L}}] + A_2^{\left(2\right)}[\vec{d}_{2,1}^{\mathrm{L}}\left(\varrho\right)\cdot\vec{E}_{\omega}^{\mathrm{L}}][\vec{d}_{1,0}^{\mathrm{L}}\left(\varrho\right)\cdot\vec{E}_{\omega}^{\mathrm{L}}].\label{eq:a2},
\end{equation}
and an analogous expression for $a_1(\varrho)$. The expected value of the permanent electric quadrupole operator in
the molecular frame $\langle Q_{p,q}^{\mathrm{M}}\left(\varrho\right)\rangle$
$\equiv$ $\langle\Psi^{\mathrm{M}}(\vec{r}^{\mathrm{M}},\varrho)\vert$
$Q_{p,q}^{\mathrm{M}}$ $\vert\Psi^{\mathrm{M}}\left(\vec{r}^{\mathrm{M}},\varrho\right)\rangle$, where $p,q=x,y,z$,
will have a zero-frequency component of the form
\begin{equation}
\langle Q_{p,q}^{\mathrm{M}}\left(\varrho\right)\rangle_{\omega=0} %&
=\sum_{i=0}^{2}\left|a_{i}\left(\varrho\right)\right|^{2}\langle Q_{p,q}^{\mathrm{M}}\rangle_{i,i}\label{eq:Q_m_allstates}%\\
%  & =\left|a_{2}\left(\varrho\right)\right|^{2}\langle Q_{p,q}^{\mathrm{M}}\rangle_{2,2},
%\label{eq:Q_m}
\end{equation}
where $\langle Q_{p,q}^{\mathrm{M}}\rangle_{i,i}\equiv\langle\psi_{i}^{\mathrm{M}}\vert Q_{q,p}^{\mathrm{M}}\vert\psi_{i}^{\mathrm{M}}\rangle$.
%and, for the sake of simplicity, we ignored contributions from states
%$\vert 0\rangle$ and $\vert 1\rangle$. 
Transforming $\langle Q_{p,q}^{\mathrm{M}}\left(\varrho\right)\rangle_{\omega=0}$
to the laboratory frame and averaging over all molecular orientations
yields the permanent quadrupole% (see Appendix) 
\begin{equation}
\langle Q_{p,q}^{\mathrm{L}}\rangle_{\omega=0} \equiv \int\mathrm{d}\varrho \langle Q_{p,q}^{\mathrm{L}}(\varrho)\rangle_{\omega=0}.
\label{eq:Q_total}
\end{equation}
The contribution of state $|2\rangle$ to this expression reads as (see Appendix)
\begin{align}
\langle (Q_2^{\mathrm{L}})_{p,q}\rangle_{\omega=0} & \equiv\int\mathrm{d}\varrho\,\left|a_{2}\left(\varrho\right)\right|^{2}\langle Q_{p,q}^{\mathrm{L}}(\varrho)\rangle_{2,2}\label{eq:Q_unsolved}\\
 & =\langle (Q^{\mathrm{L}}_2)_{p,q}\rangle_{\omega=0}^{\left(\mathrm{achiral}\right)}+\left[A_2^{\left(1\right)*}A_2^{\left(2\right)}\chi_2^{\prime}Z_{p,q}^{\prime\mathrm{L}}+\mathrm{c.c.}\right],\label{eq:Q_solved}
\end{align}
% \begin{align}
% \langle Q_{p,q}^{\mathrm{L}}\rangle_{\omega=0} & \equiv\int\mathrm{d}\varrho\,\left|a_{2}\left(\varrho\right)\right|^{2}\langle Q_{p,q}^{\mathrm{L}}(\varrho)\rangle_{2,2}\label{eq:Q_unsolved}\\
%  & =\langle Q_{p,q}^{\mathrm{L}}\rangle_{\omega=0}^{\left(\mathrm{achiral}\right)}+\left[i\chi^{\prime}Z_{p,q}^{\prime\mathrm{L}}+\mathrm{c.c.}\right],\label{eq:Q_solved}
% \end{align}
where $\langle (Q^{\mathrm{L}}_2)_{p,q}\rangle_{\omega=0}^{\left(\mathrm{achiral}\right)}$
results from the diagonal terms in $\vert a_{2}\left(\varrho\right)\vert^{2}$
and is not enantiosensitive. $\chi_2^{\prime}$ is a
rotationally invariant molecular pseudoscalar (zero for achiral molecules) encoding the enantiosensitivity
of $\langle (Q^{\mathrm{L}}_2)_{p,q}\rangle_{\omega=0}$ and defined according to 
\begin{equation}
\chi_i^{\prime}\equiv\frac{1}{30}\left\{ \left[\left(\vec{d}_{1,0}^{\mathrm{M}}\times\vec{d}_{2,0}^{\mathrm{M}}\right)\cdot\left(\langle Q^{\mathrm{M}}\rangle_{i,i}\vec{d}_{2,1}^{\mathrm{M}}\right)\right]+\left[\left(\vec{d}_{2,1}^{\mathrm{M}}\times\vec{d}_{2,0}^{\mathrm{M}}\right)\cdot\left(\langle Q^{\mathrm{M}}\rangle_{i,i}\vec{d}_{1,0}^{\mathrm{M}}\right)\right]\right\} ,\label{eq:chi_m_Q}
\end{equation}
with $\langle Q^{\mathrm{M}}\rangle_{i,i}$ a quadrupole matrix, i.e.
$\langle Q^{\mathrm{M}}\rangle_{i,i}\vec{d}_{2,1}^{\mathrm{M}}$ and
$\langle Q^{\mathrm{M}}\rangle_{i,i}\vec{d}_{1,0}^{\mathrm{M}}$ denote
multiplications of a matrix and a vector. 
%$\chi_i^{\prime}$ is zero
%for achiral molecules, while   
$Z_{p,q}^{\prime\mathrm{L}}$ is a symmetric
field pseudotensor of rank 2. It encodes the dependence of $\langle (Q^{\mathrm{L}}_2)_{p,q}\rangle_{\omega=0}$
on the field polarization according to
\begin{equation}
Z_{p,q}^{\prime\mathrm{L}}\equiv\left(\vec{E}_{\omega}^{\mathrm{L}}\times\vec{E}_{2\omega}^{\mathrm{L}*}\right)_{p}\left(\vec{E}_{\omega}^{\mathrm{L}}\right)_{q}+\left(\vec{E}_{\omega}^{\mathrm{L}}\times\vec{E}_{2\omega}^{\mathrm{L}*}\right)_{q}\left(\vec{E}_{\omega}^{\mathrm{L}}\right)_{p}.\label{eq:Z_l_Q}
\end{equation}
This expression shows that all components of $Z_{p,q}^{\prime\mathrm{L}}$ vanish
if $\omega$ and $2\omega$ are linearly polarized parallel to each
other, or if $\omega$ and $2\omega$ are circularly polarized and
counter-rotating.

\begin{figure}[b]
\begin{centering}
\includegraphics[scale=0.25]{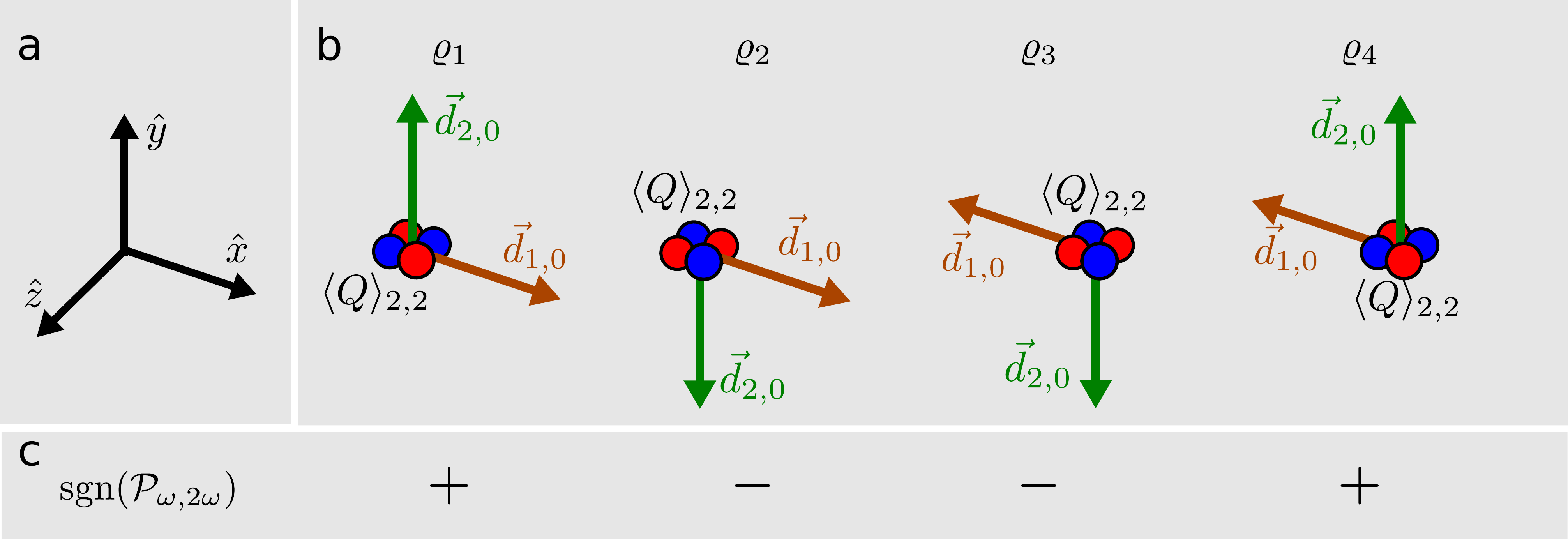}
\par\end{centering}
\caption{Simple analysis of the mechanism leading to an enantiosensitive permanent
quadrupole for a field (\ref{eq:E_xy-w2w}) and a dummy molecule with
$\vec{d}_{1,0}$, $\vec{d}_{2,0}$ , and $\langle Q\rangle_{2,2}$
oriented as shown with respect to each other and $\vec{d}_{2,1}=\vec{d}_{1,0}$.
Blue and red balls stand for negative and positive charges. \textbf{a.
}Laboratory frame. \textbf{b}. Molecular orientations with $\vec{d}_{1,0}$
aligned along $\vec{x}$ and $\vec{d}_{2,0}$ aligned along $\vec{y}$.
\textbf{c.} Sign of the interference term (\ref{eq:Pw3w}) for each
molecular orientation. The interference causes the molecular axis
$\vec{d}_{2,0}$ to become oriented, which together with the alignment
of $\vec{d}_{1,0}$ along $\vec{x}$ explains the non-vanishing permanent
quadrupole. }
\label{fig:Simple-picture_Q}
\end{figure}

For example, if we take $\omega$ and $2\omega$ linearly polarized
perpendicular to each other, say $\vec{E}_{\omega}^{\mathrm{L}}=\vec{x}^{\mathrm{L}}$
and $\vec{E}_{2\omega}^{\mathrm{L}}=e^{-i\phi}\vec{y}^{\mathrm{L}}$, 
then 
\begin{equation}
\vec{E}^{\mathrm{L}}\left(t\right)=2F\left(t\right)\left[\cos\left(\omega t\right)\vec{x}^{\mathrm{L}}+\cos\left(2\omega t+\phi\right)\vec{y}^{\mathrm{L}}\right],\label{eq:E_xy-w2w}
\end{equation}
and we obtain 
%the expected value of the electric quadrupole induced in the initially
%isotropic sample yields
\begin{equation}
\langle (Q^{\mathrm{L}}_2)_{p,q}\rangle_{\omega=0}=\langle (Q^{\mathrm{L}}_2)_{p,q}\rangle_{\omega=0}^{\left(\mathrm{achiral}\right)}+2\chi_2^{\prime}\Re\left\{A_2^{\left(1\right)*}A_2^{\left(2\right)}e^{i\phi}\right\}\left(\delta_{p,z}\delta_{q,x} + \delta_{q,z}\delta_{p,x}\right).
\end{equation}
Furthermore, one can show that for the field (\ref{eq:E_xy-w2w})
the achiral terms vanish for $p\neq q$ (see Appendix) and therefore
the enantiosensitive $xz$ component reads as
\begin{equation}
\langle (Q^{\mathrm{L}}_2)_{x,z}\rangle_{\omega=0}=2\chi_2^{\prime}\Re\left\{A_2^{\left(1\right)*}A_2^{\left(2\right)}e^{i\phi}\right\},
\label{eq:Q_2_solved}
\end{equation}
i.e., it doesn't have an achiral background and can be controlled
through the relative phase $\phi$. The other non-diagonal components
$xy$ and $yz$ vanish.

The contribution from state $\vert1\rangle$ to the permanent quadrupole (\ref{eq:Q_total}) has the same structure as Eq. (\ref{eq:Q_2_solved}), although with different coupling constants and molecular pseudoscalar $\chi_1^\prime$ [see Eq. (\ref{eq:chi_m_Q})]. Since $\vert a_0\vert^2 = 1 - \vert a_1 \vert^2 - \vert a_2 \vert^2$, the contribution from the ground state involves the coupling constants associated to $\vert1\rangle$ and $\vert2\rangle$, and a molecular pseudoscalar $\chi_0^\prime$ [see Eq. (\ref{eq:chi_m_Q})]. Together, these contributions yield
\begin{equation}
    \langle Q_{x,z}^{\mathrm{L}}\rangle_{\omega=0} = 2\left[\left(\chi_{1}^{\prime}-\chi_{0}^{\prime}\right)\Re\left\{ A_{1}^{\left(1\right)}A_{1}^{\left(2\right)*}e^{i\phi}\right\} +\left(\chi_{2}^{\prime}-\chi_{0}^{\prime}\right)\Re\left\{ A_{2}^{\left(1\right)*}A_{2}^{\left(2\right)}e^{i\phi}\right\} \right],
\end{equation}
where $A_1^{(1)}$ and $A_1^{(2)}$ are the coupling coefficients for the transitions $|0\rangle \stackrel{\omega}{\rightarrow} |1\rangle$ and $|0\rangle \stackrel{2\omega}{\rightarrow} |2\rangle \stackrel{-\omega}{\rightarrow}|1\rangle$, respectively. The other non-diagonal elements of the permanent quadrupole vanish and the diagonal terms are not enantiosensitive. 

As in the previous section, in the absence of an intermediate resonance through the state $\vert 1 \rangle$, the contribution from the second-order term in Eq. (\ref{eq:a2}) turns into a sum over all intermediate states $\vert j\rangle$. The intermediate states retain no population at the end of the pulse and the permanent quadrupole takes the form 
\begin{align}
    \langle Q_{x,z}^{\mathrm{L}}\rangle_{\omega=0} &= 2\sum_{j} (\chi_{2;j}^{\prime}-\chi_{0;j}^{\prime}) \Re\left\{A_{2}^{\left(1\right)*}A_{2;j}^{\left(2\right)}e^{i\phi}\right\}\nonumber\\
    &= 2 F_{0}^{3} \left[2\pi\delta_{\sigma}(\Delta)\right]^{2} \sum_j\frac{\chi_{2;j}^{\prime}-\chi_{0;j}^{\prime}}{\omega_{j,0}-\omega_{L}}\cos\phi,
\end{align}
where $\Delta\equiv\omega_{2,0}-2\omega_{L}$, $\chi_{i;j}$ is given by Eq. (\ref{eq:chi_m_Q}) with the replacement $1\rightarrow j$ and the other symbols were introduced as in Eq. (\ref{eq:d_total_solved_offres}).

A simple analysis analogous to that presented in
Fig. \ref{fig:Simple-picture} is shown in
Fig. \ref{fig:Simple-picture_Q} for the case of the field
(\ref{eq:E_xy-w2w}) interacting with a dummy molecule with
$\vec{d}_{1,0}^{\mathrm{M}}$, $\vec{d}_{2,0}^{\mathrm{M}}$, and
$\langle Q^{\mathrm{M}}\rangle_{2,2}$ oriented as shown and with
$\vec{d}_{2,1}^{\mathrm{M}}=\vec{d}_{1,0}^{\mathrm{M}}$.  The
population of state $\vert 2 \rangle$ is determined by the
interference term
\begin{equation}
\mathcal{P}_{\omega,2\omega}\left(\varrho\right)\equiv[\vec{d}_{2,0}\left(\varrho\right)\cdot\vec{y}][\vec{d}_{1,0}\left(\varrho\right)\cdot\vec{x}]^{2},\label{eq:Pw2w}
\end{equation}
which is positive for orientations $\varrho_{1}$ and $\varrho_{4}$ and
negative for orientations $\varrho_{2}$ and $\varrho_{3}$. This causes
the molecular axis $\vec{d}_{2,0}^{\mathrm{M}}$ to become oriented. If
$\vec{d}_{2,2}^{\mathrm{M}}$ has a non-zero component along
$\vec{d}_{2,0}^{\mathrm{M}}$, then a permanent dipole
emerges. However, this permanent dipole is contained in the
polarization plane and will therefore not change upon reflection of
the system across the polarization plane. Since this reflection is
equivalent to a change of the enantiomer, the permanent dipole is not
enantiosensitive. In contrast and as can be seen from
Fig. \ref{fig:Simple-picture_Q}, the imbalance between orientations
$\varrho_{1}$ and $\varrho_{4}$ in comparison to orientations
$\varrho_{2}$ and $\varrho_{3}$ is enough to produce an
enantiosensitive permanent quadrupole that does change upon reflection
in the polarization plane.

\section{Conclusions}

We have shown that permanent dipoles and quadrupoles can be induced
in initially isotropic samples of chiral molecules using perturbative
two-color fields that resonantly excite electronic transitions. These
permanent multipoles are enantiosensitive and their sign can be controlled
through the relative phase between the two colors. The mechanism leading
to these permanent dipoles (or quadrupoles) 
stems from uniaxial orientation of the
molecule, which occurs due to the selectivity of the 
excitation to the orientation of the molecule. Such orienting excitation can be accomplished using fields where the fundamental
and its second (or third) harmonic are linearly polarized perpendicular
to each other. The enantiosensitive permanent dipole is obtained via
three-$\omega$- vs. one-$3\omega$-photon interference. The enantiosensitive
quadrupole is obtained via two-$\omega$ vs. one-$2\omega$ interference.
In the latter case, a permanent dipole can also be generated but it
is not enantiosensitive. We expect 
these permanent multipoles to survive
for at least a few picoseconds before decaying due to molecular rotation.
Such picosecond variation of the multipoles should in principle  
lead to broadband THz emission with an enantiosensitive phase. 

Although we focused on a mechanism relying on interference between two pathways, it is also possible to induce permanent dipoles via direct pathways by relying on transitions where the photon order matters. This can be achieved e.g. using the pulse sequence in Ref. \cite{gershnabel_orienting_2018}. 

Efficient generation of enantio-sensitive permanent dipoles 
and quadrupoles 
via  orientation-sensitive excitations is possible in strong
laser fields using efficient excitation of Rydberg states via the so-called Freeman resonances \cite{freeman_above-threshold_1987} in the regime when the pronderomotive potential is comparable to the laser frequency. 
Since Rydberg states have large polarizability, we expect  significant contrast in the orientation of left and right enantiomers. Opposite orientation of left and right enantiomers and their respective induced permanent dipoles create opportunities for  enantio separation using static electric fields.

\begin{acknowledgement}
  %\verb|acknowledgement|
  We gratefully acknowledge support from the DFG SPP 1840 ``Quantum Dynamics in Tailored Intense Fields'' within the project  SM 292/5-1;
\end{acknowledgement}

\section*{Appendix}
\addcontentsline{toc}{section}{Appendix}
\subsection*{Coupling coefficients $A_f^{(n)}$}
Consider a Hamiltonian $H=H_{0}+H^{\prime}(t)$, where $H_0$ is the time-independent field-free Hamiltonian and $H^\prime(t)$ can be treated as a perturbation. If at time $t=0$ the system is in the state $\vert 0 \rangle$, the probability amplitude of finding the system in the state $\vert f \rangle$ at the time $t=T$ can be written as $a_f=a_f^{(1)} + a_f^{(2)} + \dots$, where 
\begin{align}
a_{f}^{\left(N\right)} & = \left(\frac{1}{i}\right)^{N}\int_{0}^{T}\mathrm{d}t_{N}\dots\int_{0}^{t_{3}}\mathrm{d}t_{2}\int_{0}^{t_{2}}\mathrm{d}t_{1}\langle f\vert H_{I}^{\prime}\left(t_{N}\right)\dots H_{I}^{\prime}\left(t_{2}\right)H_{I}^{\prime}\left(t_{1}\right)\vert 0\rangle
% & ={i}^{N}\sum_{j_{1},j_{2},\dots,j_{N-1}}\int_{0}^{t}\mathrm{d}t_{N}\dots\int_{0}^{t_{3}}\mathrm{d}t_{2}\int_{0}^{t_{2}}\mathrm{d}t_{1}\left[\vec{d}_{i,j_{N-1}}\cdot\vec{E}\left(t_{N}\right)e^{i\omega_{i,j_{N-1}}t_{N}}\right]\dots\left[\vec{d}_{j_{2},j_{1}}\cdot\vec{E}\left(t_{2}\right)e^{i\omega_{j_{2},j_{1}}t_{2}}\right]\left[\vec{d}_{j_{1},0}\cdot\vec{E}\left(t_{1}\right)e^{i\omega_{j_{1},0}t_{1}}\right]
\end{align}
and $H_{I}^{\prime}(t)=e^{iH_{0}t}H^{\prime}(t)e^{-iH_{0}t}$.
In the electric dipole approximation we have $H^\prime = -\vec{d}\cdot\vec{E}(t)$. For a field $\vec{E}\left(t\right)=F(t)\vec{E}_{\omega}e^{-i\omega t}+\mathrm{c.c.}$,
the contributions to $a_f^{(N)}$ from absorption of $N$ photons yield
% we get $H^{\prime}(t)\equiv H^{\prime}_{+}(t)+H^{\prime}_{-}(t)$,  $H^{\prime}_{+}(t)\equiv-(\vec{d}\cdot \vec{E}_{\omega}) F(t)e^{-i\omega t}$,  $H^{\prime}_{-}(t)\equiv H^{\prime \dagger}_{+}(t)$, where $H^{\prime}_{+}(t)$ describes absorption and $H^{\prime}_{-}(t)$ emission of a photon. In the presence of resonances only one of these two terms is important for a given transition. For example, in the case of 
\begin{equation}
a_{f}^{\left(N\right)}=\sum_{j_{1},j_{2},\dots,j_{N-1}}a_{f;j_{1},j_{2},\dots,j_{N-1}}^{\left(N\right)},
\label{eq:a_N_sum}
\end{equation}
where the sum is over the different quantum pathways through the intermediate states $\vert {j_1} \rangle$, $\vert {j_2}\rangle$, ..., $\vert {j_{N-1}}\rangle$. The amplitude of each pathway can be written as 
\begin{align}
a_{f;j_{1},j_{2}\dots,j_{N-1}}^{\left(N\right)} & =A_{f;j_{1},j_{2},\dots,j_{N-1}}^{\left(N\right)}\left(\omega\right)(\vec{d}_{f,j_{N-1}}\cdot\vec{E}_{\omega})\dots (\vec{d}_{j_{2},j_{1}}\cdot\vec{E}_{\omega})(\vec{d}_{j_{1},0}\cdot\vec{E}_{\omega}),
\label{eq:a_N_pathway}
\end{align}
The coupling coefficient $A_{f;j_{1},j_{2},\dots,j_{N-1}}^{\left(N\right)}(\omega)$ carries the information about the frequency of the light, its envelope, and the detunings according to
\begin{multline}
A_{f;j_{1},j_{2},\dots,j_{N-1}}^{(N)}(\omega)=i^{N}\int_{0}^{T}\mathrm{d}t_{N} F\left(t_{N}\right)e^{i\left(\omega_{f,j_{N-1}}-\omega\right)t_{N}}\dots \\ \times \int_{0}^{t_{3}}\mathrm{d}t_{2} F\left(t_{2}\right)e^{i\left(\omega_{j_{2},j_{1}}-\omega\right)t_{2}} \int_{0}^{t_{2}}\mathrm{d}t_{1} F\left(t_{1}\right)e^{i\left(\omega_{j_{1},0}-\omega\right)t_{1}},
\label{eq:A_N}
\end{multline}
where $\omega_{ij} \equiv \omega_i - \omega_j$. Contributions to $a_f^{(N)}$ from pathways involving photon emissions require exchanging $\vec{E}_\omega$ by $\vec{E}_\omega^*$ in Eq. (\ref{eq:a_N_pathway})\footnote{If the transition dipoles are complex then one must also complex conjugate them. Here we assume they are real.} and $\omega$ by $-\omega$ in Eq. (\ref{eq:A_N}) in the corresponding transitions.

In the case of a resonant pathway the sum in Eq. (\ref{eq:a_N_sum}) reduces to a single term, which is the assumption in several parts of the main text. There we write $A_f^{(N)}$ as a shorthand for $A_{f;j_{1},j_{2},\dots,j_{N-1}}^{(N)}$.

\subsection*{Orientation integrals required in Sec. \ref{sec:quadrupole}}
Replacing Eq. (\ref{eq:a2}) in Eq. (\ref{eq:Q_unsolved}) we obtain 
\begin{align}
\langle Q_{p,q}^{\mathrm{L}}\rangle_{\omega=0} & =\vert A_2^{\left(1\right)}\vert^{2}I_{p,q}^{\left(2\omega\right)}+\vert A_2^{\left(2\right)}\vert^{2}I_{p,q}^{\left(\omega\right)}+\left[A_2^{\left(1\right)*}A_2^{\left(2\right)}I_{p,q}^{\left(\omega,2\omega\right)}+\mathrm{c.c.}\right],\label{eq:Q_expanded}
\end{align}
where the integrals $I_{p,q}^{\left(2\omega\right)}$, $I_{p,q}^{\left(\omega\right)}$,
and $I_{p,q}^{\left(\omega,2\omega\right)}$ are defined by
\begin{equation}
I_{p,q}^{\left(2\omega\right)}\equiv\int\mathrm{d}\varrho\left|\vec{d}_{2,0}^{\mathrm{L}}\left(\varrho\right)\cdot\vec{E}_{2\omega}^{\mathrm{L}}\right|{}^{2}\langle Q_{p,q}^{\mathrm{L}}\rangle_{2,2},\label{eq:I_2w}
\end{equation}
\begin{equation}
I_{p,q}^{\left(\omega\right)}\equiv\int\mathrm{d}\varrho\left|[\vec{d}_{2,1}^{\mathrm{L}}\left(\varrho\right)\cdot\vec{E}_{\omega}^{\mathrm{L}}][\vec{d}_{1,0}^{\mathrm{L}}\left(\varrho\right)\cdot\vec{E}_{\omega}^{\mathrm{L}}]\right|^{2}\langle Q_{p,q}^{\mathrm{L}}\rangle_{2,2},\label{eq:I_w}
\end{equation}
\begin{equation}
I_{p,q}^{\left(\omega,2\omega\right)}\equiv\int\mathrm{d}\varrho\,[\vec{d}_{2,1}^{\mathrm{L}}\left(\varrho\right)\cdot\vec{E}_{\omega}^{\mathrm{L}}][\vec{d}_{1,0}^{\mathrm{L}}\left(\varrho\right)\cdot\vec{E}_{\omega}^{\mathrm{L}}][\vec{d}_{2,0}^{\mathrm{L}}\left(\varrho\right)\cdot\vec{E}_{2\omega}^{\mathrm{L}*}]\langle Q_{p,q}^{\mathrm{L}}\rangle_{2,2}.\label{eq:I_w2w}
\end{equation}
These integrals can be solved following the procedure in Ref. \cite{andrews_threedimensional_1977}.
We will first solve $I_{p,q}^{\left(\omega,2\omega\right)}$ for arbitrary
polarizations and then show that $I_{2\omega}$ and $I_{\omega}$
vanish when $p\neq q$, $\vec{E}_{\omega}^{\mathrm{L}}=\vec{x}$,
and $\vec{E}_{2\omega}^{\mathrm{L}}=e^{i\phi}\vec{y}^{\mathrm{L}}$.

\subsubsection*{$\boldsymbol{I_{p,q}^{\left(\omega,2\omega\right)}}$ \label{subsec:Iw2w}}

We are dealing with an integral of the form 
\begin{align}
I_{i_{4}i_{5}} & =\int\mathrm{d}\varrho\left(\vec{a}^{\mathrm{L}}\cdot\vec{B}^{\mathrm{L}}\right)\left(\vec{b}^{\mathbb{\mathrm{L}}}\cdot\vec{B}^{\mathrm{L}}\right)\left(\vec{c}^{\mathrm{L}}\cdot\vec{C}^{\mathrm{L}}\right)Q_{i_{4},i_{5}}^{\mathrm{L}}\nonumber \\
 & =I_{i_{1}i_{2}i_{3}i_{4}i_{5};\lambda_{1}\lambda_{2}\lambda_{3}\lambda_{4}\lambda_{5}}^{\left(5\right)}a_{\lambda_{1}}^{\mathrm{M}}b_{\lambda_{2}}^{\mathrm{M}}c_{\lambda_{3}}^{\mathrm{M}}Q_{\lambda_{4},\lambda_{5}}^{\mathrm{M}}B_{i_{1}}^{\mathrm{L}}B_{i_{2}}^{\mathrm{L}}C_{i_{3}}^{\mathrm{L}},
\end{align}
where
\begin{equation}
I_{i_{1}i_{2}i_{3}i_{4}i_{5};\lambda_{1}\lambda_{2}\lambda_{3}\lambda_{4}\lambda_{5}}^{\left(5\right)}\equiv\int\mathrm{d}\varrho l_{i_{1}\lambda_{1}}l_{i_{2}\lambda_{2}}l_{i_{3}\lambda_{3}}l_{i_{4}\lambda_{4}}l_{i_{5}\lambda_{5}}
\end{equation}
$\vec{a}^{\mathrm{M}}$, $\vec{b}^{\mathrm{M}}$, and $\vec{c}^{\mathrm{M}}$
are arbitrary vectors fixed in the molecular frame, and $Q_{i_{4},i_{5}}^{\mathrm{M}}$
is an arbitary symmetric second-rank tensor fixed in the molecular
frame. The transformation to the laboratory frame is given by $v_{i}^{\mathrm{L}}\left(\varrho\right)=l_{i\lambda}\left(\varrho\right)v_{\lambda}^{\mathrm{M}}$
for vectors and $Q_{i_{1},i_{2}}^{\mathrm{L}}\left(\varrho\right)=l_{i_{1}\lambda_{1}}\left(\varrho\right)l_{i_{2}\lambda_{2}}\left(\varrho\right)Q_{\lambda_{1},\lambda_{2}}^{\mathrm{M}}$
for the second-rank tensor, where $l_{i\lambda}\left(\varrho\right)$
is the matrix of direction cosines, we sum over repeated indices and
use latin indices for components in the laboratory frame and greek
indices for components in the molecular frame. $\vec{B}^{\mathrm{L}}$
and $\vec{C}^{\mathrm{L}}$ are arbitrary vectors fixed in the laboratory
frame. Using Eq. (31) in Ref. \cite{andrews_threedimensional_1977}
we obtain
\global\long\def\f#1#2#3#4#5{\epsilon_{\lambda_{#1}\lambda_{#2}\lambda_{#3}}\delta_{\lambda_{#4}\lambda_{#5}}\epsilon_{i_{#1}i_{#2}i_{#3}}\delta_{i_{#4}i_{#5}}}%
\begin{align}
I_{i_{4}i_{5}} & =\frac{1}{30}\bigg[\f 13425+\f 13524+\f 23415\nonumber \\
 & +\f 23514\bigg]a_{\lambda_{1}}^{\mathrm{M}}b_{\lambda_{2}}^{\mathrm{M}}c_{\lambda_{3}}^{\mathrm{M}}Q_{\lambda_{4},\lambda_{5}}^{\mathrm{M}}B_{i_{1}}^{\mathrm{L}}B_{i_{2}}^{\mathrm{L}}C_{i_{3}}^{\mathrm{L}}
\end{align}
where we used $\epsilon_{i_{1}i_{2}i_{3}}B_{i_{2}}B_{i_{3}}=\epsilon_{\lambda_{1}\lambda_{2}\lambda_{3}}Q_{\lambda_{2}\lambda_{3}}=0$.
The first term can be rewritten as 
\begin{align}
  \f 13425 & a_{\lambda_{1}}^{\mathrm{M}}b_{\lambda_{2}}^{\mathrm{M}}c_{\lambda_{3}}^{\mathrm{M}}Q_{\lambda_{4},\lambda_{5}}^{\mathrm{M}}B_{i_{1}}^{\mathrm{L}}B_{i_{2}}^{\mathrm{L}}C_{i_{3}}^{\mathrm{L}} \nonumber\\
  & =\left(\vec{a}^{\mathrm{M}}\times\vec{c}^{\mathrm{M}}\right)_{\lambda_{4}}Q_{\lambda_{4},\lambda_{5}}^{\mathrm{M}}b_{\lambda_{5}}^{\mathrm{M}}\left(\vec{B}^{\mathrm{L}}\times\vec{C}^{\mathrm{L}}\right)_{i_{4}}B_{i_{5}}^{\mathrm{L}}\nonumber \\
 & =\left[\left(\vec{a}^{\mathrm{M}}\times\vec{c}^{\mathrm{M}}\right)\cdot\left(Q^{\mathrm{M}}\vec{b}^{\mathrm{M}}\right)\right]\left(\vec{B}^{\mathrm{L}}\times\vec{C}^{\mathrm{L}}\right)_{i_{4}}B_{i_{5}}^{\mathrm{L}}.
\end{align}
Analogous operations for the rest of the terms yield
\begin{align}
  I_{i_{4}i_{5}}&=\frac{1}{30}\left\{ \left[\left(\vec{a}^{\mathrm{M}}\times\vec{c}^{\mathrm{M}}\right)\cdot\left(Q^{\mathrm{M}}\vec{b}^{\mathrm{M}}\right)\right]+\left[\left(\vec{b}^{\mathrm{M}}\times\vec{c}^{\mathrm{M}}\right)\cdot\left(Q^{\mathrm{M}}\vec{a}^{\mathrm{M}}\right)\right]\right\} \nonumber\\
  & \times \left\{ \left(\vec{B}^{\mathrm{L}}\times\vec{C}^{\mathrm{L}}\right)_{i_{4}}B_{i_{5}}^{\mathrm{L}}+\left(\vec{B}^{\mathrm{L}}\times\vec{C}^{\mathrm{L}}\right)_{i_{5}}B_{i_{4}}^{\mathrm{L}}\right\} \label{eq:I_w2w_solved}
\end{align}
Performing the substitutions $\{\vec{a},\vec{b},\vec{c},Q\}\rightarrow\{\vec{d}_{2,1},\vec{d}_{1,0},\vec{d}_{2,0},\langle Q\rangle_{2,2}\}$,
$\{\vec{B},\vec{C}\}\rightarrow\{\vec{E}_{\omega},\vec{E}_{2\omega}^{*}\}$,
and $\{i_{4},i_{5}\}\rightarrow\{p,q\}$ and using Eqs. (\ref{eq:Q_expanded})
and (\ref{eq:I_w2w}) yields Eqs. (\ref{eq:Q_solved})-(\ref{eq:Z_l_Q}).

\subsubsection*{$\boldsymbol{I_{p,q}^{\left(2\omega\right)}}$}

Assuming a linearly polarized $\vec{E}_{2\omega}$ we must deal with
an integral of the form 
\begin{align}
I_{i_{3}i_{4}} & =\int\mathrm{d}\varrho\,[\vec{a}^{\mathrm{L}}\cdot\vec{B}^{\mathrm{L}}][\vec{a}^{\mathrm{L}}\cdot\vec{B}^{\mathrm{L}}]Q_{i_{3}i_{4}}^{\mathrm{L}}\nonumber\\
 & =I_{i_{1}i_{2}i_{3}i_{4};\lambda_{1}\lambda_{2}\lambda_{3}\lambda_{4}}^{\left(4\right)}a_{\lambda_{1}}^{\mathrm{M}}a_{\lambda_{2}}^{\mathrm{M}}Q_{\lambda_{3},\lambda_{4}}^{\mathrm{M}}B_{i_{1}}^{\mathrm{L}}B_{i_{2}}^{\mathrm{L}},
\end{align}
where 
\begin{equation}
I_{i_{1}i_{2}i_{3}i_{4};\lambda_{1}\lambda_{2}\lambda_{3}\lambda_{4}}^{\left(4\right)}\equiv\int\mathrm{d}\varrho l_{i_{1}\lambda_{1}}l_{i_{2}\lambda_{2}}l_{i_{3}\lambda_{3}}l_{i_{4}\lambda_{4}},
\end{equation}
and we use the same notation as in the previous subsection. Using
Eq. (19) in Ref. \cite{andrews_threedimensional_1977} we get
\begin{equation}
I_{i_{3}i_{4}}=\vec{F}_{i_{3}i_{4}}^{\left(4\right)}\cdot M^{\left(4\right)}\vec{G}_{i_{3}i_{4}}^{\left(4\right)},
\end{equation}
where $\vec{F}_{i_{3}i_{4}}^{\left(4\right)}$ is given by 
\begin{equation}
\vec{F}_{i_{3}i_{4}}^{\left(4\right)}=\left(\begin{array}{c}
\delta_{i_{1}i_{2}}\delta_{i_{3}i_{4}}\\
\delta_{i_{1}i_{3}}\delta_{i_{2}i_{4}}\\
\delta_{i_{1}i_{4}}\delta_{i_{2}i_{3}}
\end{array}\right)B_{i_{1}}^{\mathrm{L}}B_{i_{2}}^{\mathrm{L}}=\left(\begin{array}{c}
\left|\vec{B}^{\mathrm{L}}\right|^{2}\delta_{i_{3}i_{4}}\\
B_{i_{3}}^{\mathrm{L}}B_{i_{4}}^{\mathrm{L}}\\
B_{i_{3}}^{\mathrm{L}}B_{i_{4}}^{\mathrm{L}}
\end{array}\right)
\end{equation}
For $\vec{B}^{\mathrm{L}}=\vec{y}^{\mathrm{L}}$ we have $B_{i}^{\mathrm{L}}=\delta_{iy}$
and therefore $B_{i_{3}}^{\mathrm{L}}B_{i_{4}}^{\mathrm{L}}=B_{i_{3}}^{\mathrm{L}}B_{i_{4}}^{\mathrm{L}}=\delta_{i_{3}y}\delta_{i_{4}y}=\delta_{i_{3}i_{4}}\delta_{i_{3}y}$,
which yields $\vec{F}_{i_{3}i_{4}}^{(4)}\propto\delta_{i_{3}i_{4}}$ and
$I_{i_{3}i_{4}}\propto\delta_{i_{3}i_{4}}$. The substitutions $\{\vec{a},Q,\vec{B},i_{3},i_{4}\}\rightarrow\{\vec{d}_{2,0},\langle Q\rangle_{2,2},\vec{y},p,q\}$
then yield $I_{p,q}^{\left(2\omega\right)}\propto\delta_{p,q}$.

\subsubsection*{$\boldsymbol{I_{p,q}^{\left(\omega\right)}}$}

Assuming a linearly polarized $\vec{E}_{\omega}$ we must deal with
an integral of the form 
\begin{align}
I_{i_{5},i_{6}} & =\int\mathrm{d}\varrho[\vec{a}^{\mathrm{L}}\cdot\vec{B}^{\mathrm{L}}][\vec{b}^{\mathrm{L}}\cdot\vec{B}^{\mathrm{L}}][\vec{a}^{\mathrm{L}}\cdot\vec{B}^{\mathrm{L}}][\vec{b}^{\mathrm{L}}\cdot\vec{B}^{\mathrm{L}}]Q_{i_{5}i_{6}}^{\mathrm{L}}\\
 & =I_{i_{1}i_{2}i_{3}i_{4}i_{5}i_{6};\lambda_{1}\lambda_{2}\lambda_{3}\lambda_{4}\lambda_{5}\lambda_{6}}^{\left(6\right)}a_{\lambda_{1}}^{\mathrm{M}}b_{\lambda_{2}}^{\mathrm{M}}a_{\lambda_{3}}^{\mathrm{M}}b_{\lambda_{4}}^{\mathrm{M}}Q_{\lambda_{5},\lambda_{6}}^{\mathrm{M}}B_{i_{1}}^{\mathrm{L}}B_{i_{2}}^{\mathrm{L}}B_{i_{3}}^{\mathrm{L}}B_{i_{4}}^{\mathrm{L}},
\end{align}
where 
\begin{equation}
I_{i_{1}i_{2}i_{3}i_{4}i_{5}i_{6};\lambda_{1}\lambda_{2}\lambda_{3}\lambda_{4}\lambda_{5}\lambda_{6}}^{\left(6\right)}=\int\mathrm{d}\varrho l_{i_{1}\lambda_{1}}l_{i_{2}\lambda_{2}}l_{i_{3}\lambda_{3}}l_{i_{4}\lambda_{4}}l_{i_{5}\lambda_{5}}l_{i_{6}\lambda_{6}},
\end{equation}
and we use the same notation as in the previous subsections. Using
Tab. II in \cite{andrews_threedimensional_1977} we have that 
\begin{equation}
I_{i_{5},i_{6}}=\vec{F}_{i_{5}i_{6}}^{\left(6\right)}\cdot M^{\left(6\right)}\vec{G}_{i_{3}i_{4}}^{\left(6\right)},
\end{equation}
where $\left(F_{i_{5}i_{6}}^{\left(6\right)}\right)_{r}\equiv f_{r}^{\left(6\right)}B_{i_{1}}^{\mathrm{L}}B_{i_{2}}^{\mathrm{L}}B_{i_{3}}^{\mathrm{L}}B_{i_{4}}^{\mathrm{L}}$
and $f_{r}^{\left(6\right)}$ ($r=1,2,\dots,15$) is given in Tab.
II in \cite{andrews_threedimensional_1977}. For $\vec{B}^{\mathrm{L}}=\vec{x}^{\mathrm{L}}$
we have $B_{i}^{\mathrm{L}}=\delta_{ix}$ and therefore 
\begin{equation}
\left(F_{i_{5}i_{6}}^{\left(6\right)}\right)_{r}=\begin{cases}
\delta_{i_{5}i_{6}}, & r=1,4,7\\
\delta_{i_{5}x}\delta_{i_{6}x}, & \mathrm{otherwise}
\end{cases}
\end{equation}
Since $\delta_{i_{5}x}\delta_{i_{6}x}=\delta_{i_{5}i_{6}}\delta_{i_{5}x}$,
then $\vec{F}_{i_{5}i_{6}}^{\left(6\right)}\propto\delta_{i_{5}i_{6}}$
and $I_{i_{5},i_{6}}\propto\delta_{i_{5}i_{6}}$. The substitutions
$\{\vec{a},\vec{b},Q,\vec{B},i_{5},i_{6}\}\rightarrow\{\vec{d}_{1,0},\vec{d}_{2,1},\langle Q\rangle_{2,2},\vec{x},p,q\}$
then yield $I_{p,q}^{\left(\omega\right)}\propto\delta_{p,q}$.

\bibliographystyle{spphys}
\bibliography{MyLibrary}
\end{document}